Comments on
six papers published by  S.P. Anjali Devi and R. Kandasamy  in
Heat and Mass Transfer, ZAMM, Mechanics Research Communications,
International Communications in Heat and Mass Transfer,
Communications in Numerical Methods in Engineering, Journal of
Computational and Applied Mechanics

Asterios Pantokratoras
Associate Professor of Fluid Mechanics
School of Engineering, Democritus University of Thrace,
67100 Xanthi – Greece
e-mail:apantokr@civil.duth.gr

Comment on
"Effects of chemical reaction, heat and mass transfer on laminar flow along a semi infinite horizontal plate", by S.P. Anjali Devi and R. Kandasamy [**Heat and Mass Transfer**, 35, 1999, pp. 465-467]

In the above paper an analysis has been carried out to obtain results in a steady laminar boundary layer flow over a motionless horizontal plate placed in a horizontal free stream. The plate temperature is constant and different from the free stream temperature. In addition it is assumed that mass transfer takes place along the plate accompanied by a chemical reaction of this substance. The boundary layer equations are transformed into ordinary ones and subsequently are solved using the Runge-Kutta Gill method. However, there are some errors in this paper which are presented below:

1. In nomenclature the symbol $U_0$ is denoted as "velocity of plate" but in the present problem the plate is motionless.

2. In page 465 it is mentioned that "the present investigation deals with the study of flow of a fluid in the presence of chemical reaction with a species concentration due to the MHD flow". However the above work has no relation to Magnetohydrodynamics (MHD).

3. In page 466 it is mentioned that "values are obtained for Prandtl number Pr=0.7" whereas in figures 2 and 3 the Prandtl number is 1.

4. In figures 2 and 3 a new parameter, denoted by M, appears that has not defined in the paper.

5. In the transformed energy (temperature) equation (16) the symbol n (different from η which is the transverse coordinate) appears. This quantity is denoted in the nomenclature as "rate of chemical reaction" but has not defined in the paper. If we suppose that this quantity really exists and is relevant to chemical reaction then it must be included in the mass transfer equation (17) where the chemical reaction takes place. In the temperature equation (16) no chemical reaction takes place and the inclusion of n in this equation is unreasonable.

6. The dimensionless velocity f˝(η) should vary between 0 and 1 and the dimensionless concentration φ(η) should vary between 1 and 0 according to boundary conditions given in equation 18. However this does not happen in figures 2 and 3 and these figures are wrong

Comment on
"Effects of chemical reaction, heat and mass transfer on MHD flow past a semi infinite plate", by S.P. Anjali Devi and R. Kandasamy [**ZAMM**, 80, 2000, pp. 697-701]

In the above paper a numerical solution for the steady MHD flow over an infinite horizontal plate in the presence of species concentration and chemical reaction has been presented. The boundary layer equations are transformed into ordinary ones and subsequently are solved using the Runge-Kutta Gill method. However, there are some errors in this paper which are presented below:

1. In nomenclature the symbol n is denoted as "parameter associated with thermal stratification". This symbol appears in equation (18) and takes the value 0.25 in page 699. However in the present problem no thermal stratification exists. The ambient temperature is constant and equal to $T_\infty$.

2. The momentum equation used by the authors is

$$u\frac{\partial u}{\partial x} + v\frac{\partial u}{\partial y} = \nu\frac{\partial^2 u}{\partial y^2} + g\beta(T - T_\infty) + g\beta^*(C - C_\infty) - \frac{\sigma B_0^2}{\rho}u \qquad (1)$$

where u and v are the velocity components, $\nu$ is the fluid kinematic viscosity, $\rho$ is the fluid density, $\sigma$ is the fluid electrical conductivity, $B_0$ is the strength of the magnetic field, g is the gravity acceleration, T is the fluid temperature, C is the species concentration, $\beta$ is the fluid thermal expansion coefficient and $\beta^*$ is the species expansion coefficient. The boundary conditions are:

at y = 0:   u =v=0 , T=$T_w$ , C=$C_w$        (2)

as y $\rightarrow \infty$  u = $U_\infty$ , T=$T_\infty$, C=$C_\infty$        (3)

Let us apply the momentum equation at large y. At large distances from the plate the fluid temperature is equal to ambient temperature and the buoyancy term $g\beta(T-T_\infty)$ in momentum equation is zero. The same happens with the term $g\beta^*(C-C_\infty)$.

Taking into account that, at large distances from the plate, velocity is everywhere constant and equal to $U_\infty$ the velocity gradient $\partial u/\partial y$ is also zero. The same happens with the diffusion term $v\partial^2 u/\partial y^2$. This means that the momentum equation takes the following form at large y

$$U_\infty \frac{\partial U_\infty}{\partial x} = -\frac{\sigma B_0{}^2}{\rho} U_\infty \qquad (4)$$

or

$$\frac{\partial U_\infty}{\partial x} = -\frac{\sigma B_0{}^2}{\rho} \qquad (5)$$

From the above equation we see that the free stream velocity should change along x and therefore the momentum equation (1) is not compatible with the assumption that the free stream velocity is constant. Therefore the momentum equation is wrong.

3. The dimensionless velocity f″(η) should vary between 0 and 1 and the dimensionless temperature θ(η) and the dimensionless concentration φ(η) should vary between 1 and 0 according to boundary conditions (equation 20 in their work). However this does not happen in figures 2-5 and all these figures are wrong.

## Comment on

### "Thermal stratification effects  on laminar boundary-layer flow over a wedge with suction or injection", by S.P. Anjali Devi and R. Kandasamy
### [**Mechanics Research Communications**, 28, 2001, pp. 349-354]

In the above paper an analysis has been carried out to obtain results in a steady laminar boundary layer flow along a wedge with suction or injection at the wedge surface. The wedge  temperature is constant whereas the  free stream temperature increases along the wedge (thermal stratification). The boundary layer equations are transformed into ordinary ones and subsequently are solved using the Runge-Kutta Gill  method.

The boundary conditions given in equation (24) are:

at $\eta = 0$:    $f''(0)=0$ , $\theta(0) =1$          (1)

as $\eta \rightarrow \infty$    $f'(\infty)=1$ , $\theta(\infty)=0$          (2)

Taking into account the above boundary conditions the dimensionless velocity $f'(\eta)$ should vary between 0 and 1 and  the dimensionless temperature $\theta(\eta)$  should vary  between 1 and 0. However in figure 3 the temperature $\theta(\eta)$     varies between   -0.10 and 0 and the transverse coordinate $\eta$ between 0 and 0.9. Someone may argue  that there is a confusion between the axes and the variable of the horizontal axis, which varies between 0 and 0.9, represents the dimensionless temperature and the vertical axis represents the transverse coordinate $\eta$. However in figure 3 the variable in the vertical axis is negative and the transverse coordinate $\eta$ can not take negative values (it varies between 0 and $\infty$ according to above equations).  This problem exists in  figures 2, 3, 4 and 5  and all these figures are  wrong.

Comment on
"Effects of chemical reaction, heat and mass transfer on non-linear MHD laminar boundary-layer flow over a wedge with suction or injection", by S.P. Anjali Devi and R. Kandasamy [**International Communications in Heat and Mass Transfer**, 29, 2002, pp. 707-716]

In the above paper an analysis has been carried out to obtain results in a steady MHD laminar boundary layer flow over a wedge with suction or injection. The wedge  temperature is constant and different from the free stream temperature. In addition it is assumed that mass transfer takes place along the wedge accompanied by  a chemical reaction of this substance. The boundary layer equations are transformed into ordinary ones and subsequently are solved using the Runge-Kutta Gill  method.

The boundary conditions given in equation (30) are:

at $\eta = 0$:      $f'(0)=0$, $\theta(0)=1$, $\varphi(0)=1$                 (1)

as $\eta \to \infty$    $f'(\infty)=1$ , $\theta(\infty)=0$,  $\varphi(\infty)=0$         (2)

Taking into account the above boundary conditions the dimensionless velocity $f'(\eta)$ should vary between 0 and 1,  the dimensionless temperature $\theta(\eta)$ should vary  between 1 and 0 and the dimensionless species concentration $\varphi(\eta)$ should vary  between 1 and 0.  However in figure 3 the temperature $\theta(\eta)$   varies between  -4.2 and 0 and  the transverse coordinate $\eta$ between 0 and 1. Someone may argue  that there is a confusion between the axes and the variable of the horizontal axis, which varies between 0 and 1, represents the dimensionless temperature and the vertical axis represents the transverse coordinate $\eta$. However in figure 3 the variable in the vertical axis is negative and the transverse coordinate $\eta$ can not take negative values (it varies between 0 and $\infty$ according to above equations).  This problem exists in  figures 2-10  and all these figures are wrong.

Comment on
"Effects of chemical reaction, heat and mass transfer on non-linear MHD flow over an accelerating surface with heat source and thermal stratification in the presence of suction or injection" by S. P. Anjali Devi and R. Kandasamy [**Communications in Numerical Methods in Engineering**, 19, 2003, pp. 513-520]

In the above paper a numerical solution for the steady laminar MHD boundary layer flow over a vertical plate with suction or injection in the presence of mass diffusion has been obtained. The plate temperature and the ambient temperature vary along the plate. The species concentration at the plate varies along the plate whereas the ambient concentration is constant. The boundary layer equations are transformed into ordinary ones and subsequently are solved using the Runge-Kutta Gill method. However, there are some errors in this paper which are presented below:

1. The velocity of the vertical plate is linear according to law $u=\alpha x$ (equation 5) whereas the dimensionless quantities (Reynolds number, Grashof number, chemical reaction parameter) are defined using a reference velocity U that has not been defined in the paper.

2. In figures 1, 2 and 3 a parameter $r_2$ is used which has not been defined in the paper. Apparently this is the quantity r that is used in the other figures and is defined in equation (5).

3. The dimensionless velocity $f'(\eta)$, the dimensionless temperature $\theta(\eta)$ and the dimensionless species concentration $\varphi(\eta)$ should vary between 1 and 0 according to boundary conditions given in equation 20. However this does not happen and the figures 1-6 are wrong.

Comment on
"Effects of chemical reaction, heat and mass transfer on non-linear laminar boundary-layer flow over a wedge with suction or injection", by R. Kandasamy and S.P. Anjali Devi [**Journal of Computational and Applied Mechanics**, 5, 2004, pp. 21-31]

In the above paper the boundary layer flow along a wedge is investigated. The plate temperature is constant and different from the ambient temperature. In addition it is assumed that mass transfer takes place along the wedge accompanied by a chemical reaction of this substance. The substance concentration at the wedge surface is constant and different from that of the ambient stream. The boundary layer equations are transformed into ordinary ones and subsequently are solved using the Runge-Kutta Gill method. However, there are some errors in this paper which are presented below:

1. The Reynolds number defined in equation (12) is wrong.
2. In figure 2 the dimensionless velocity $f''(\eta)$ changes from 0 to 25 and the transverse coordinate $\eta$ from 0 to 1. Apparently this is wrong. The dimensionless velocity should change from 0 to 1 and the transverse coordinate from 0 to 25. The same happens with all figures and therefore all figures are wrong. Someone may argue that there is a confusion between the axes and $f''(\eta)$, $\theta(\eta)$ and $\varphi(\eta)$ are shown on the horizontal axis. However in figure 4, the variable shown in the horizontal axis, which may represent the species concentration, does not vary between 0 and 1.
3. In all figures a parameter Ec, which takes the value 0.001 (Ec=0.001), appears. However this parameter has not been defined in the paper. Usually this symbol denotes the Eckert number when the viscous dissipation term is included in the energy equation. However, in the present problem no viscous dissipation term exists.
4. In conclusions it is mentioned that " due to the uniform magnetic field, in the case of suction" but in the present problem no magnetic field exists.

In conclusion all the above papers are of very low quality, written without care and are partly or completely wrong.